\documentclass[twocolumn,showpacs,preprintnumbers,amsmath,amsfonts,amssymb,floatfix,aps,pra,superscriptaddress]{revtex4}
\usepackage{graphicx}
\usepackage{amssymb}
\usepackage{amsmath}
\usepackage{amsfonts}
\usepackage{bm}
\usepackage{color}
\usepackage{float}
\usepackage{tikz} 
\usepackage[com pat=1.1.0]{tikz-feynman}
\usepackage[nodisplayskipstretch]{setspace}
\usetikzlibrary{arrows.meta,decorations.markings}
\usetikzlibrary{bending} 
\tikzset{
    arc arrow/.style args={%
    to pos #1 with length #2}{
    decoration={
        markings,
         mark=at position 0 with {\pgfextra{%
         \pgfmathsetmacro{\tmpArrowTime}{#2/(\pgfdecoratedpathlength)}
         \xdef\tmpArrowTime{\tmpArrowTime}}},
        mark=at position {#1-\tmpArrowTime} with {\coordinate(@1);},
        mark=at position {#1-2*\tmpArrowTime/3} with {\coordinate(@2);},
        mark=at position {#1-\tmpArrowTime/3} with {\coordinate(@3);},
        mark=at position {#1} with {\coordinate(@4);
        \draw[-{Latex[length=#2,bend]}]       
        (@1) .. controls (@2) and (@3) .. (@4);},
        },
     postaction=decorate,
     }
}
\usepackage{filecontents}
\usepackage{tikzscale}
\usepackage{hyperref}

\newcommand{\be}{\begin{equation}}
\newcommand{\ee}{\end{equation}}

\begin{document}

\title{Mobile impurity probing  a two-dimensional  superfluid phase transition }
\author{R. Alhyder}%
\email{Corresponding author: ragheed.alhyder@phys.au.dk}
\affiliation{Center for Complex Quantum Systems, Department of Physics and Astronomy, Aarhus University, Ny Munkegade 120, DK-8000 Aarhus C, Denmark
}%
\author{G. M. Bruun}%
\affiliation{Center for Complex Quantum Systems, Department of Physics and Astronomy, Aarhus University, Ny Munkegade 120, DK-8000 Aarhus C, Denmark
}%
\affiliation{Shenzhen Institute for Quantum Science and Engineering and Department of Physics, Southern University of Science and Technology, Shenzhen 518055, China
}%

\begin{abstract}
The use of atomically sized quantum systems as highly sensitive measuring devices represents an exciting and quickly growing research field. Here, we explore the properties of a quasiparticle formed by a mobile impurity interacting with a two-dimensional fermionic superfluid. The energy of the quasiparticle is shown to be lowered by superfluid pairing as
this increases the compressibility of the Fermi gas, thereby  making it easier for the impurity to perturb its surroundings. We demonstrate that the fundamentally 
  discontinuous nature of the superfluid to normal phase transition of a two-dimensional system, leads to a rapid increase in the quasiparticle energy around 
  the critical temperature. The magnitude of this increase  exhibits a nonmonotonic behavior as a function of the pairing strength with a sizable maximum in the 
  cross-over region, where the spatial extend of the Cooper pairs is comparable to the interparticle spacing.  
 Since the quasiparticle energy is measurable with present experimental techniques, our results illustrate how   impurities entangled with their environment can serve as useful  probes for non-trivial thermal and quantum correlations. 
\end{abstract}

\maketitle

\section{Introduction}
The realisation of accurate measuring devices, which are based on their quantum mechanical properties represents a new and exciting  research
  direction with great technological potential. A main goal is to develop atomically sized probes with maximal sensitivity and minimal back-action 
   on the environment~\cite{Degen2017}. 
Impurity atoms are  promising candidates for this,  and they have  already been used experimentally to measure the 
temperature~\cite{Olf:2015un,Hohmann2016,Bouton2020} and density~\cite{adam2021coherent}  of a surrounding 
quantum degenerate  gas, as well as to  detect induced interactions~\cite{Edri2020}.
So far, the vast majority of investigations into mobile impurities in atomic gases  concerned cases where  the environment 
is either a weakly interacting Bose-Einstein condensate (BEC) or an ideal Fermi gas, 
and the impurity forms a quasiparticle called the Bose or Fermi polaron respectively~\cite{jorgensen2016,hu2016bose,Ardila2019,Yan2020,schirotzek2009ofp,kohstall2011metastability,Koschorreck2012,Scazza2017,Fritsche2021}.
 
Much less attention has been payed to  mobile impurities in environments with  correlations between the particles. 
The properties of an impurity in a fermionic superfluid across the strongly correlated
 BCS-BEC cross-over were examined~\cite{Nishida2015,Yi2015}, but a general description turns out to be complicated by the presence of
   ultraviolet divergencies related to three-body physics~\cite{Pierce2019}. A particularly 
 interesting case concerns two-dimensional (2D) systems, where quantum and thermal fluctuations are more pronounced than in 3D and  true long range order is prohibited at a non-zero temperature~\cite{mermin1966absence,Hohenberg1967}. Nevertheless, 2D systems can exhibit a so-called Berezinskii-Kosterlitz-Thouless (BKT) phase transition to a 
 superfluid phase with quasi-long range order~\cite{Berezinskii1972,Kosterlitz1972,Kosterlitz1973}, which has been observed in a range of 
 bosonic systems including  
 $^4$He films~\cite{Bishop1978}, magnetic layers~\cite{Durr1989}, and  atomic Bose gases~\cite{Hadzibabic2006,Clade2009,Tung2010,Hung2011,Desbuquois:2012tz}.
A 2D superfluid  in a two-component  strongly interacting atomic fermi gas was observed only recently~\cite{Ries2015,Murthy2015,Sobirey2021}, 
and the underlying discontinuity of the phase transition has so far not been seen unambiguously in this system. In general, 
 our understanding of 2D fermionic superfluids is   less developed as compared to their bosonic counterparts.
 
Here, we investigate a mobile impurity immersed in  a 2D fermionic superfluid. Interactions between the impurity and the surrounding fermions lead to the formation 
of a quasiparticle, i.e.\ a polaron, and we show that its energy is lowered due to an increase in the compressibility of the environment caused by superfluid pairing. We furthermore demonstrate that the abrupt vanishing of  the superfluid density  at the critical temperature, characteristic for a  superfluid to normal phase transition in 2D~\cite{Nelson1977}, gives
 rise to a rapid increase in the polaron energy. The increase depends   non-monotonically on the Fermi-Fermi interaction strength exhibiting a 
 maximum when the size of the Cooper pairs is comparable to the interparticle spacing. Our results show how a mobile impurity can serve as a sensitive probe for  
  thermal and quantum correlations  of a 2D fermionic system thereby providing important guides for improving 
  our understanding of its non-trivial  superfluid to normal phase transition.

\begin{figure}
    \centering
  \includegraphics[width=\columnwidth]{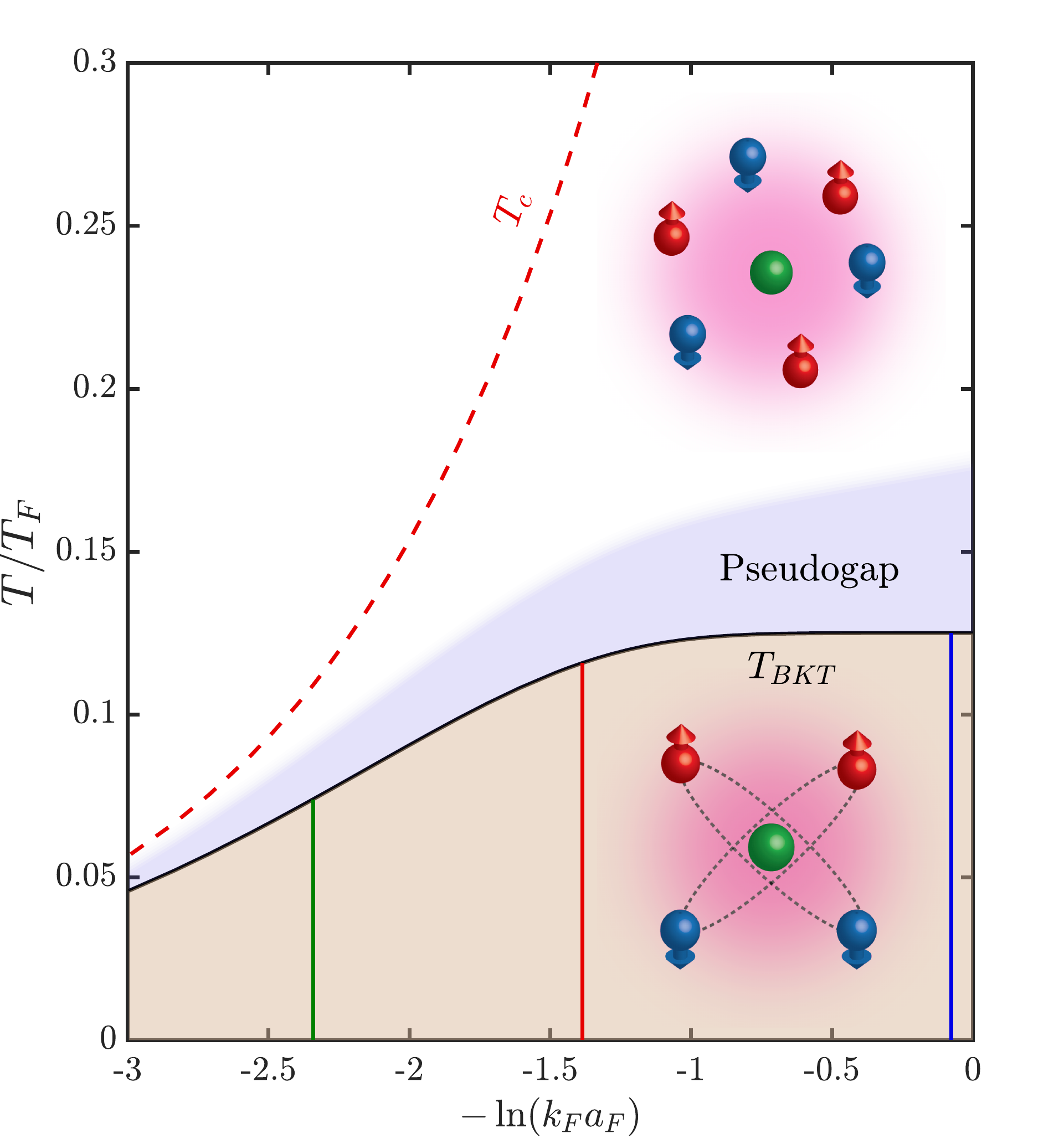}
    \caption{We consider a mobile impurity (green ball without arrow) forming a  quasiparticle by interacting with a two-component Fermi gas  (blue and red balls with arrows) in 2D. An attractive 
    interaction with strength $-\ln (k_Fa_F)$ between the fermions gives rise to a discontinuous phase transition between a superfluid 
    and a normal phase at the critical temperature $T_\text{BKT}$, which is suppressed from the mean-field BCS prediction $T_c$ by phase fluctuations. 
    The vertical lines indicate the coupling strengths for which we plot the polaron energy in Fig.~\eqref{PolaronEnergyFig}. 
    } 
    \label{Phasediagram}
\end{figure}

\section{System}
Consider an impurity of mass $m_I$ immersed in a two-component ($\sigma=\uparrow, \downarrow$) gas of  fermions of mass $m$ in 2D.  The $\uparrow$ and $\downarrow$
 fermions interact attractively and form a superfluid below a critical temperature. 
 Using BCS theory to describe  this superfluid phase, the Hamiltonian of the system is
\begin{align}
	\hat{H} &= \sum_{\bm{k}\sigma} \xi_{\bm{k}} \hat{a}^\dagger_{\bm{k}\sigma} \hat{a}_{\bm{k}\sigma}  + \Delta\sum_{\bm{k}} (\hat{a}_{-\bm{k}\downarrow} \hat{a}_{\bm{k}\uparrow}+\hat{a}^\dagger_{-\bm{k}\uparrow} \hat{a}^\dagger_{\bm{k}\downarrow} )\nonumber\\
	&+ \sum_{\bm{k}} \epsilon_{\bm{k}} \hat{c}^\dagger_{\bm{k}} \hat{c}_{\bm{k}} +
	g_{\text{IF}}\!\sum_{\bm{k}\bm{k}'\bm{q}\sigma}\hat{a}^\dagger_{\bm{k}'-\bm{q}\sigma} \hat{a}_{\bm{k}'\sigma}\hat{c}^\dagger_{\bm{k}+\bm{q}} \hat{c}_{\bm{k}}.
	\label{eq:Hamiltonian}
\end{align}
Here, $ \hat{a}^\dagger_{\bm{k}\sigma} $ is the creation operator of a fermion with momentum $\bm{k}$, spin $\sigma$,  and kinetic energy $\xi_{\bm{k}} =  k^2/2m-\mu$ with $\mu$ the chemical potential, 
and $ \hat{c}_{\bm{k}} $ creates an impurity with momentum $\bm{k}$ and kinetic energy $\epsilon_{\bm{k}}= k^2/2m_I$,
$g_{\text{IF}}$ is the interaction strength for momenta less than a cutoff $\Lambda$, it can be eliminated in favor of a 2-body bound state with energy $\epsilon_B$ using \cite{Randeria1990, Zollner2011}:
\begin{align}
\frac{1}{g_{\text{IF}}} = -\frac{1}{\mathcal{V}}\sum_{|\bm{q}|<\Lambda} \frac{1}{E_B+q^2/2m_r}
\label{g_0}
\end{align}
where $m_r=mm_i/(m+m_i)$ and $E_B=1/2m_ra_{IF}^2$.\\
The superfluid gap $\Delta$ at temperature $T$ is determined from 
\begin{align}
\begin{split}
	\Delta & = -\frac{g_{\text{F}}}{\beta \mathcal{V}}\sum_{\bm{k},n}G_{12}(\bm{k},i\omega_n)
		\end{split}
	\end{align}
		where $G_{12}$ is the anomalous Green's function. 
This leads to the following gap equation
\begin{equation}
 \int dk\, k \Bigg(\frac{\tanh(E_{\bm{k}}/2T )}{2E_{\bm{k}}}-\frac{1}{\epsilon_B+k^2/2m_r}\Bigg)=0,
 \label{Gapeqn}
\end{equation}
where $E_{\bm{k}} = \sqrt{\xi_{\bm{k}}^2+\Delta^2}$
and we have renormalised the gap equation by replacing the Fermi-Fermi interaction strength by the energy of a bound state of two fermions $\epsilon_B$, which is always present for an attractive interaction \cite{Randeria1990}. We work in units where $\hbar$, $k_B$, and the system volume are all unity.  

\section{The BKT transition}
The 2D superfluid with quasi-long range order melts into a normal phase when vortex and anti-vortex pairs unbind and proliferate. This occurs at the critical temperature 
determined by the 
condition~\cite{Kosterlitz1972,kosterlitz1974critical}
\begin{align}
T_\text{BKT} =\frac{\pi}{8m}n_s(T_\text{BKT})
\label{KTCondition}
\end{align}
where $n_s$ is the superfluid density given by~\cite{lifshitz2013statistical}
 \begin{align}
\begin{split}
\frac{n_s(T)}{n}&=1+\frac{1}{2 \pi m n}\int_0^\infty \! \text{d}k  k^3 \frac{\partial f(E_{\bm{k}})}{\partial E_{\bm k}}.
\label{SuperfluidDen}
\end{split}
\end{align}
Here $f(E)=[\exp(E/T)+1]^{-1}$ is the Fermi-Dirac distribution and $n=k_F^2/2\pi$ is the total density from both spin states of the Fermi gas. 
The integral is always negative and vanishes at $T=0$ ensuring that the superfluid density is always equal to or smaller than the total density. 
It follows from Eq.~\eqref{KTCondition} that the superfluid density of the Fermi gas  exhibits a universal jump   $\Delta n_s/m T_\text{BKT}=8/\pi$
at the BKT transition. This jump  has not yet 
 been observed in the experiments exploring two-dimensional  atomic fermi gases, and a main goal here is to demonstrate that the  discontinuity of the 
 phase transition  can be detected by looking at the properties of the impurity.

Fig.~\eqref{Phasediagram} shows the phase diagram of the Fermi gas as a function of the Fermi-Fermi interaction length 
strength parametrised by $-\ln(k_Fa_F)$ and temperature. Here, $a_F$ is a scattering length defined by writing the energy of the bound Fermi-Fermi
 dimer  as $-1/ma_F^2$. 
 The Fermi gas is in a superfluid phase below a critical temperature $T_\text{BKT}$ obtained by solving Eqs.~\eqref{Gapeqn}-\eqref{SuperfluidDen}
 self-consistently. We vary the chemical potential $\mu$ to keep the  density 
 $n=2\sum_{\mathbf k}[v_{\mathbf k}^2(1-f_{\mathbf k})+u_{\mathbf k}^2f_{\mathbf k}]$ fixed, where 
  $v_{\bm{k}}^2=1-u_{\bm{k}}^2=(1-\xi_{\bm{k}}/E_{\bm{k}})/2$ are the  coherence factors.

 For weak coupling $-\ln(k_Fa_F)\ll -1$ corresponding to the so-called BCS regime with large Cooper pairs, 
  the superfluid transition temperature  is close to 
that obtained from mean-field BCS theory, which predicts a smooth decrease of the superfluid density to zero at the critical temperature $T_c$. 
It follows that the jump in the superfluid density at $T_\text{BKT}$  is small for weak coupling.
For stronger coupling however, phase fluctuations significantly suppress the critical temperature below the BCS prediction leading to a large 
jump in the superfluid density at the phase transition. We obtain $T_{\text{BKT}} = T_F/8$ in the strong coupling regime with $-\ln(k_Fa_F)\gtrsim -1$ 
reflecting that the superfluid density equals the total density $n=k_F^2/2\pi$ just below the transition giving rise to a maximal jump $\Delta n_s=n$ ~\cite{Babaev1999,Botelho2006,Salasnich2013}. 
We note however that the gas  eventually enters the  BEC regime with increasing $-\ln(k_Fa_F)\gtrsim 1$, where 
it can be described as a Bose gas of  tightly 
bound Cooper pairs with a BKT critical temperature that decreases slowly~\cite{fisher1988,Prokofev2001,Ries2015,Murthy2015}.

\section{Perturbation theory}
We now turn to the properties of the impurity in the  Fermi gas. Since the  case of general interaction strengths between the impurity and the fermions as well as 
between the fermions is very complicated, we focus on the regime of weak impurity-fermion interactions where a reliable perturbative theory 
can be developed. 

To do this, consider the scattering matrix between the impurity and a fermion. As detailed in appendix \ref{appendixA}, it can in the ladder  approximation be written as 
\begin{align}
	\mathcal{T}(k)	 = \frac{g}{1 - g\Delta\Pi(k)} \simeq g + g^2\Delta\Pi(k)+\ldots
	\label{Tperturbative}
\end{align}
where $k=(\bm{k},i\omega_n)$ denotes the center-of-mass momentum $\bm{k}$ of the colliding pair with $i\omega_n$  a Matsubara frequency,
and an expression for the pair propagator $\Delta\Pi(k)$ is given in appendix \ref{appendixB}. It follows from Eq.~\eqref{Tperturbative} that 
\begin{align}
g=-\frac{\pi}{m_r} \frac{1}{\ln (k_F a_{\text{IF}})}
\label{CouplingStrength}
\end{align} 
is  an effective 2D interaction  strength between the impurity and the fermions~\cite{Bloom1975,Engelbrecht1992}. We have thus 
eliminated the  bare impurity-Fermi coupling strength $g_\text{IF}$ in favour of an effective interaction strength, which is a function of the  
  energy $-1/2m_ra_\text{IF}^2$ of the bound impurity-fermion dimer   with $a_\text{IF}$ the impurity-fermion scattering length and 
$m_r=mm_I/(m+m_I)$ the reduced mass~\cite{Randeria1990, Zollner2011}. When $|\ln (k_F a_{\text{IF}})|\gg 1$, 
 the  effective interaction is weak  and  the impurity properties can be calculated reliably using perturbation theory in $g$ as used in Eq.~\eqref{Tperturbative}.

To first order in $g$, the energy shift of the impurity is simply given by the mean-field expression  $\Sigma_1(p)=gn$. The second order term is  
\begin{align}
\Sigma_2(p) &=
2Tg^2\sum_{k}\left[G_{11}(k)\Delta\Pi(p+k)+G_{12}(k)\Pi_{21}(p+k)\right]\nonumber\\
&=Tg^2\sum_{k}G_0(k)\chi(p-k)
\label{Selfenergy}
\end{align}
where
\begin{align}
G_{11}(\bm{k},i\omega_n)&=\frac{u_{\bm{k}}^2}{i\omega_n-E_{\bm k}}+\frac{v_{\bm{k}}^2}{i\omega_n+E_{\bm k}},\\
G_{12}(\bm{k},i\omega_n)&=u_{\bm{k}}v_{\bm{k}}\left(\frac{1}{i\omega_n+E_{\bm k}}-\frac{1}{i\omega_n-E_{\bm k}}\right)
\label{GreensFn}
\end{align}
are the normal and anomalous propagators for the superfluid  and $G_0^{-1}(k)=i\omega_n-\epsilon_k$ is the impurity 
Green's function in the absence of interactions. The first term in the first line of Eq.~\eqref{Selfenergy} describes the coupling of the impurity to particle-hole excitations in the superfluid whereas the 
second term describes coupling to pair breaking excitations. 
In the second line of Eq.~\eqref{Selfenergy}, we write the impurity self-energy in terms of the density-density correlation function $\chi(k)$ of the superfluid, 
which gives the compressibility for long wave lengths. Expressions for 
 $\Pi_{12}(k)$ and $\chi(k)$ are given in the appendix \ref{appendixB}. 

The energy $\epsilon_P$ of the polaron  can now be found from the pole of the impurity Green's function $G(k)$ analytically continued to real frequencies. Using the Dyson equation 
$G^{-1}(k)=G_0^{-1}(k)-\Sigma(k)$ with $\Sigma(k)=\Sigma_1(k)+\Sigma_2(k)$  to second order in $g$ then yields 
\begin{align}
\epsilon_P=gn+\Sigma_2(0,\epsilon_P+i0_+),
\end{align}
where we take zero  momentum and $m_I=m$ in the following for simplicity.

\section{Normal phase}
\label{normalphase}
Strictly speaking, it is only the superfluid density that exhibits a discontinuity at $T_{BKT}$. This is because the long range phase coherence of the gap  is lost for 
$T>T_{BKT}$ whereas its amplitude $|\Delta|$ remains continuous. In the following, we will nevertheless assume that the pairing gap jumps to zero at the temperature 
$T_{BKT}$ for the BKT phase transition. That is, we will use BCS theory for $T\le T_{BKT}$ and set $\Delta=0$ for $T> T_{BKT}$, corresponds to assuming that the Fermi gas is non-interacting. 
The reasons for this are the following. First, vortices with a vanishing  gap in their centers proliferate at the transition temperature, which will significantly 
decrease the average gap for  $T>T_{BKT}$. Second, even if the gap is non-zero in the normal phase it does not lead
to a perfect vanishing of the density of states around the Fermi level as opposed to in the superfluid phase. Indeed, the gap in the normal phase is often refereed to as a pseudo-gap  
for this reason, and its description requires inclusion of fluctuation effects beyond BCS theory. A central feature of the theories for the pseudo-gap region is that 
 predict a suppressed but \emph{non-zero} density of states at the 
Fermi level~\cite{Bauer2014,Marsiglio2015}. This means that impurity can scatter on low energy excitations in the surrounding bath even for a non-zero pseudo-gap in the normal phase above $T_{BKT}$.
Using a non-zero $\Delta$ in the BCS Green's functions Eq.~\eqref{GreensFn} above $T>T_{BKT}$ would on the other hand yield a vanishing density of the states at the Fermi level thereby 
missing these low energy excitations completely, which likely will result in an unphysical polaron energy.  
 Finally, BCS theory drastically overestimates the temperature for which the amplitude of the gap vanishes except for weak coupling. Indeed, more sophisticated theories predict that the pseudo-gap vanishes for temperatures much lower than the mean-field critical temperature~\cite{Bauer2014,Klimin_2012}. 
   It follows that the pseudo-gap goes to zero in a temperature range above $T_{BKT}$ that is narrow compared to the BCS transition temperature. 
Given these facts, it is physically reasonable as a first approximation to assume that the gap jumps to zero at  $T_{BKT}$, which 
should therefore be understood as the limiting form of a continuous but sharp drop.  
We therefore take $\Delta=0$ for $T>T_{BKT}$ in the rest of the paper, which corresponds to assuming that the Fermi gas is ideal in the normal phase.

\section{Results}
\begin{figure}[h]
    \centering
    \includegraphics[width=\columnwidth]{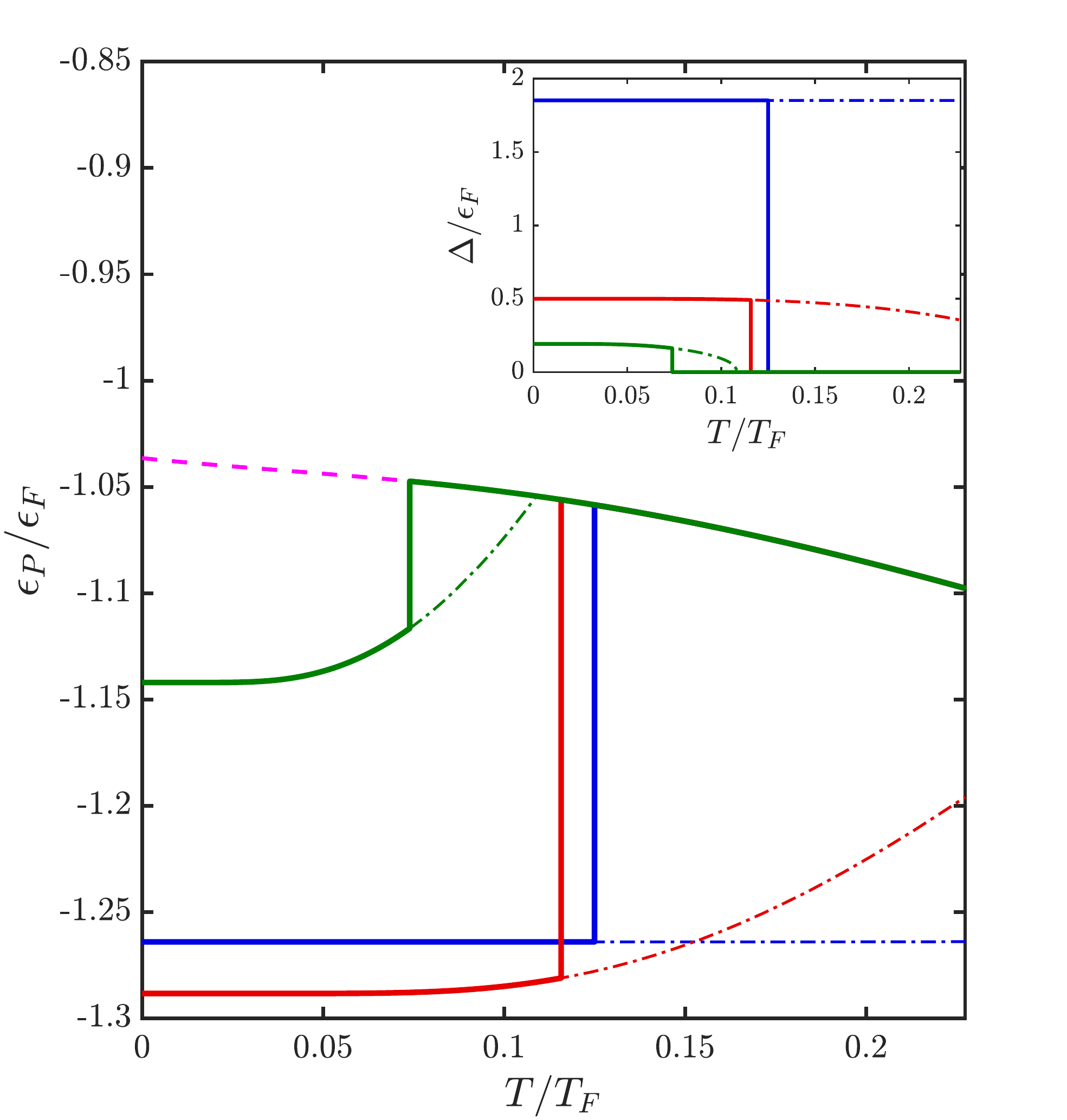}
    \caption{Polaron energy as a function of temperature for the Fermi-Fermi coupling strengths   
    $-\ln(k_Fa_F)=-2.342$ (green upper line),  $-\ln(k_Fa_F)=-1.386$ (red lower line), and $-\ln(k_Fa_F)=-0.077$ (blue middle line). The dashed pink line gives polaron energy in the normal phase, and the 
    dash-dotted lines the polaron energy assuming the Fermi gas remains superfluid up to the mean-field critical temperature $T_c$. 
    The inset shows the superfluid gap as 
    a function of temperature for the same coupling strengths with the dash-dotted lines giving the mean-field prediction for $T>T_\text{BKT}$. The colors blue (upper), red (middle), and green (lower) correspond to the scattering values of the main plot.
    The dotted lines illustrate the expected continuous behavior of the polaron energy and  magnitude $|\Delta|$ of the gap 
     around the BKT transition as explained in section \ref{normalphase}.
    } 
    \label{PolaronEnergyFig}
\end{figure}

In Fig.~\eqref{PolaronEnergyFig}, we plot the  polaron energy as a function of the temperature for different values of the Fermi-Fermi interaction, which are 
shown by vertical lines in Fig.~\eqref{Phasediagram}. 
This is found by first solving Eqs.~\eqref{Gapeqn}-\eqref{SuperfluidDen} numerically for constant density 
 to find the properties of the Fermi bath. We then calculate the  impurity self-energy from Eq.~\eqref{Selfenergy}. The impurity-fermion interaction is $-\ln (k_Fa_{\text{IF}})= -1.03$ for which 
  perturbation theory is still accurate for an impurity in a 2D ideal Fermi gas~\cite{schmidt20122dpolaron}. 
First, we note that the polaron has a lower energy when the Fermi gas is in the superfluid phase as compared to when it is in the normal phase (dashed lines). 
Physically, this is because pairing correlations increase the compressibility of the Fermi gas~\cite{Leggett1980,Mark2012,Guo2013,Tajima:2017ue} 
so that the impurity more easily can perturb its surroundings  thereby lowering the energy. 
As the temperature increases, the superfluid gap decreases (inset in Fig.~\eqref{PolaronEnergyFig}), and the polaron energy approaches the value in the normal phase, 
which in turn decreases with temperature in analogy with what is  found in 3D~\cite{Hu2018,Tajima_2018}. In particular,  the polaron energy exhibits a 
discontinuity  at the critical temperature $T_\text{BKT}$, since we assume that the superfluid gap in the surrounding medium  jumps 
to zero at the transition. It is strictly  only the superfluid density that is discontinuous at the transition 
due to the loss of phase coherence, whereas as the amplitude of the gap and therefore the polaron energy is continuous. Nevertheless, as we argued above the  
gap must be expected to exhibit a steep decrease in a narrow region around $T_\text{BKT}$. The results 
shown in Fig.~\eqref{PolaronEnergyFig} should in this sense be understood as a limiting form of  a continuous but rapid increase in $\epsilon_P$ on the scale of the mean-field (BCS) transition 
temperature. 
This should be contrasted to BCS theory  predicting a gap that  goes to zero at a much higher transition temperature $T_c$
(except for weak coupling), giving rise to a much more smooth behaviour of polaron energy. Thus, the abrupt change of the 
Fermi gas at $T_\text{BKT}$ is  reflected in the polaron energy, which exhibits a sizeable and steep increase in its energy in a narrow 
temperature region. Importantly, this should be observable with the spectral resolution of current Fermi polaron experiments \cite{schirotzek2009ofp,kohstall2011metastability,Koschorreck2012,Scazza2017,Fritsche2021}. 
These results show that the polaron can be used as a probe of  abrupt nature of the superfluid to normal phase transition in a 2D fermionic superfluid. In particular, a
measurement of the polaron energy and its behaviour around $T_\text{BKT}$ should provide valuable information regarding the 
 thermal and quantum fluctuations in the critical region.

\begin{figure}[h]
    \centering
    \includegraphics[width=\columnwidth]{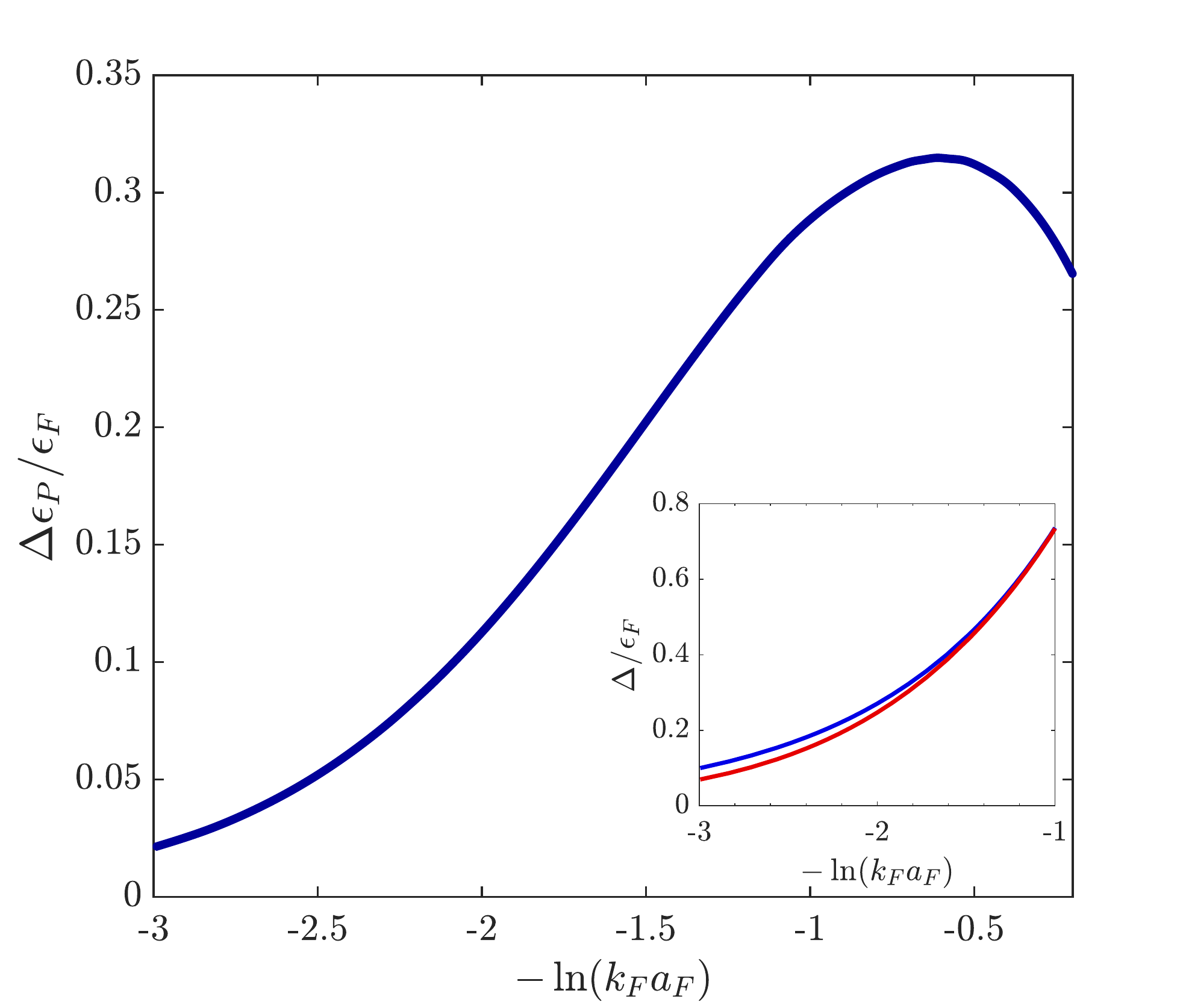}
    \caption{The limiting discontinuity $\Delta\epsilon_P$ in the polaron energy at the critical temperature $T_\text{BKT}$ as a function of the Fermi-Fermi coupling strength
    $-\ln(k_Fa_F)$. The inset shows the superfluid gap at $T=0$ (blue upper line) and at $T=T_{\text{BKT}}$ (red lower line) as a function of the interaction strength. 
   } 
   \label{PolaronJumpFig}
\end{figure}

 Figure \eqref{PolaronEnergyFig} moreover shows that the amplitude of the rapid energy increase  is larger  for $-\ln(k_Fa_F)=-1.386$ than for $-\ln(k_Fa_F)=-2.342$. This is as expected, since a stronger 
Fermi-Fermi coupling gives rise to a larger decrease  in the superfluid gap (inset) at $T_\text{BKT}$.
 The discontinuity is however smaller again for even stronger coupling with $-\ln(k_Fa_F)=-0.077$,  even though the pairing energy is larger.
To explore this  further,  we plot in  Fig.~\eqref{PolaronJumpFig}  the  value $\Delta\epsilon_P=\epsilon_P(T_\text{BKT}^+)-\epsilon_P(T_\text{BKT}^-)$ as a function of the Fermi-Fermi  
 coupling strength $-\ln(k_Fa_F)$, which 
corresponds to the change in the polaron energy in the limit where it occurs with infinite slope at $T_\text{BKT}$. We see that $\Delta\epsilon_P$ initially increases with the coupling strength in the BCS regime. It  reaches a sizeable maximum of 
 $\Delta\epsilon_P\simeq0.32\epsilon_F$ in the cross-over region around $\ln(k_Fa_F)\sim -0.6$, after which it 
  decreases  as the BEC region is approached with increasing  interaction, 
 even though the gap continues to increase as shown in the inset. This should be contrasted to the gap, which increases monotonically as a function of the coupling both for $T=0$ and
 $T=T_{BKT}$ as shown in the inset of 
 Fig.~\eqref{PolaronJumpFig}. Interestingly, the maximum in  $\Delta\epsilon_P$ occurs when the size of the Cooper pairs is comparable to the interparticle spacing, where one has also 
  observed a maximum in the 
 critical temperature~\cite{Ries2015,Murthy2015} and in the critical velocity~\cite{Sobirey2021}.
  As the coupling strength increases further with $-\ln(k_Fa_F)\gg 1$,  the Cooper pairs shrink and the Fermi gas becomes a BEC of dimers with a transition temperature in a narrow region 
   $\propto 1/\ln[\ln(1/na_D^2)]$ below the mean-field prediction with $a_D$ the dimer-dimer scattering length, and   a correspondingly small discontinuity in the 
   superfluid density~\cite{fisher1988,Prokofev2001,Ries2015,Murthy2015}. It follows that there \emph{should} be a maximum 
   in $\Delta\epsilon_P$ somewhere in the cross-over region as indeed predicted here. Note however that  our  theory 
 is unreliable in the BEC regime, since it does not 
 include the Bogoliubov-Anderson mode, which becomes the dominant excitation compared to particle-hole and pair breaking excitations. 
  A consistent description of the polaron in the whole  BEC-BCS cross-over of the Fermi gas is remains an open and very challenging problem beyond the present scope.   

\begin{figure}[h]
    \centering
    \includegraphics[width=\columnwidth]{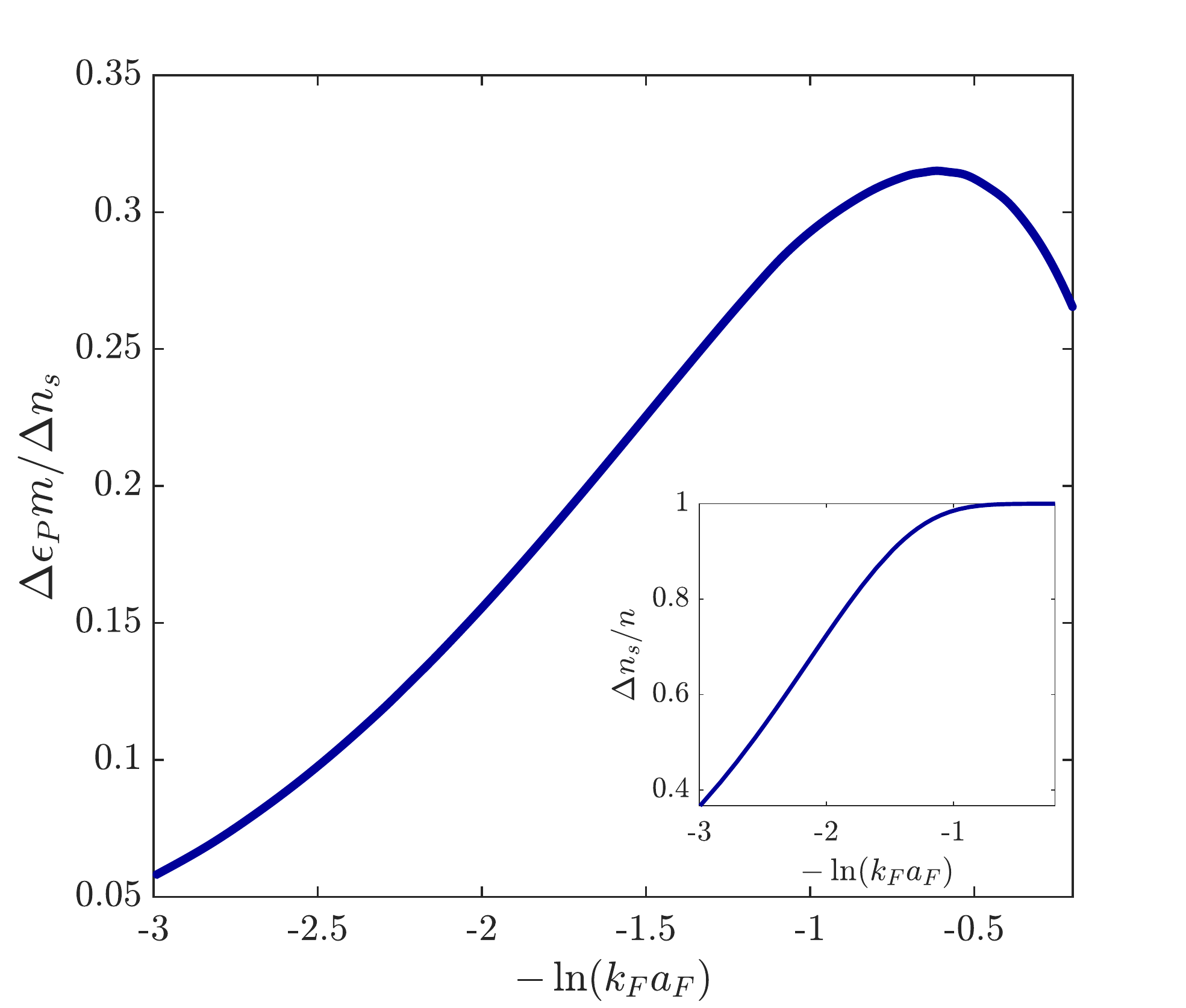}
    \caption{ The limiting discontinuity $\Delta\epsilon_P$ in the polaron energy at the critical temperature $T_\text{BKT}$ in units of the jump $\Delta n_s$ in the superfluid density
    as a function of the Fermi-Fermi coupling strength 
    $-\ln(k_Fa_F)$.  The inset shows  $\Delta n_s/n$ as a function of interaction strength. 
     } 
    \label{PolaronJumpnsFig}
\end{figure}

Finally, Fig.~\eqref{PolaronJumpnsFig} plots the limiting change $\Delta\epsilon_P$ in the polaron energy at $T_{BKT}$ 
in units of the jump $ \Delta n_s/m$ in the superfluid density at the critical temperature, which is shown in the inset. 
This shows that while the rapid change in the polaron energy is a direct consequence of the abrupt nature of the BKT phase transition of the surrounding medium, 
there is no simple proportionality between $\Delta \epsilon_P$ and $\Delta n_s$. Instead, the ratio $\Delta\epsilon_Pm/\Delta n_s$ increases with Fermi-Fermi 
interaction strength in the BCS regime  reaching a maximum at $\ln(k_Fa_F)\sim -0.6$ after which is decreases. The jump in the superfluid density at the transition temperature 
on the other hand increases monotonically with the coupling strength, reaching the limiting value $n$ in the BEC regime.

\section{Discussion and outlook}
We investigated the properties of a quasiparticle  formed by a mobile impurity in a  fermionic superfluid. The characteristic discontinuity of the 
superfluid to normal phase transition of a 2D system was shown to give rise to a rapid increase 
 in the quasiparticle energy around the transition temperature. We demonstrated that the amplitude of the increase 
 depends non-monotonically on the pairing strength with a maximum in the cross-over region between the BCS and BEC limits, and that it is measurable with present 
 experimental techniques. In particular, radio-frequency spectroscopy has proven to be a very powerful technique for measuring the polaron energy 
 both in Fermi gases and in BECs~\cite{jorgensen2016,hu2016bose,Ardila2019,Yan2020,schirotzek2009ofp,kohstall2011metastability,Koschorreck2012,Scazza2017,Fritsche2021}.

 Our results show a way to probe the properties of the  superfluid as well as its transition to the normal phase for a 2D fermionic system. This is of particular interest since there is no 
 quantitatively reliable theory for this challenging problem in the strong coupling regime. For instance, a non-zero width of the rapid increase of the polaron energy around $T_\text{BKT}$ will 
 provide information regarding the pseudogap region. This could moreover cast light on the intriguing question regarding the role of vortex-antivortex pairs in the superfluid phase and above, and their interplay with the impurity. 
It would also be very interesting  to develop an improved theoretical 
understanding regarding the properties of the BKT transition for a fermionic superfluid including its quantum and thermal fluctuations~\cite{Bauer2014,Marsiglio2015,Bighin2016,Mulkerin2017}. Here, experimental results regarding the polaron energy in the critical region would provide important guides for this challenging problem.

 The magnitude of the jump in the polaron energy will clearly be even larger for stronger interactions between the impurity and the surrounding Fermi gas. Exploring this requires 
 going beyond the perturbative approach  used here, which is an interesting topic for future study. 
Another fascinating but very challenging problem is to explore how the quasiparticle  evolves smoothly from a Fermi to a Bose polaron 
as the Fermi gas changes from a BCS superfluid to a BEC of dimers.

From a broader perspective, our results  illustrate how impurities entangled with their environment via particle collisions can be used a sensitive probes for non-trivial quantum and thermal correlations. 
This motivates further investigations into  how coherent superpositions of internal spin states of the impurity can be used to enhance the sensitivity of the impurity probe 
while minimising the back-action  on the environment~\cite{Klein_2007,Ng2008,Mehboudi_2019,Mitchison2020}. Another intriguing research direction is 
 to investigate how  impurities can be used to probe non-local correlations and order, as well as the geometric and topological properties  of the 
 environment~\cite{Grusdt:2016ul,Guardian2019,Munoz2020,Pimenov2021,Baldelli2021}.  

\acknowledgments{
This work has been supported by the Danish National Research Foundation through the Center of Excellence "CCQ" (Grant agreement no.: DNRF156), the Independent Research Fund Denmark- Natural Sciences via Grant No. DFF -8021-00233B.}
\appendix
\section{$\mathcal{T}$ matrix}
\label{appendixA}
In order to get an expression for $g_{\text{IF}}$ in terms of the pair propagator we write the $\mathcal{T}$ matrix in vacuum as
\begin{align}
\begin{split}
	\mathcal{T}_v(\bm{k},\omega) = \frac{1}{g_{\text{IF}}^{-1} - \Pi_v(\bm{k},\omega)},
	\label{Tvac}
\end{split}
\end{align}
where $\Pi_v$ is the pair propagator for two fermions in a vacuum
\begin{align}
\begin{split}
\Pi_v (\bm{k},\omega) &= 
\frac{1}{\mathcal{V}}\sum_{\bm{q}}^\Lambda
\frac{1}{\omega - k^2/2M  - q^2/2m_r },
\end{split}
\end{align}
with $M=m+m_I$ and
 $\Lambda$ is a UV cutoff that we send to infinity at the end of the calculation. We perform a variable change $q^2=x \Rightarrow dq=dx/2\sqrt{x}$ and after a straightforward calculation we get
 \begin{align}
\begin{split}
\Pi_v (\bm{k},\omega) & = -\frac{m_r}{2\pi}\Big( \ln (|\frac{\omega+i0_+ - k^2/2M -\Lambda}{\omega_n - k^2/2M + \mu}|) + i \pi \\&-  i \arg(\omega+i0_+ - k^2/2M )\Big).
\label{eq:PiV}
\raisetag{15pt}
\end{split}
\end{align}
Lastly, since the dimer bound state energy is a pole of the vacuum $\mathcal{T}$ matrix we can write from Eq. \eqref{Tvac}
$g_{\text{IF}}^{-1} = \Pi_v(0,\epsilon_B)$. 

We can now write the scattering matrix in the medium as 
\begin{align}
\begin{split}
	\mathcal{T}(\bm{k},i\omega_n) = \frac{1}{\Pi_v(0,\epsilon_B) - \Pi(\bm{k}, \omega_n)}
	\label{Tpert}
\end{split}
\end{align}
where 
\begin{align}
\begin{split}
	& \Pi(\bm{k},i\omega_n) =\frac{1}{\mathcal{V}}\sum_{\bm{p}}
\Bigg(\frac{u_{\bm{p}}^2(1-n_{\bm{p}})}{i\omega_n-E_{\bm p}-\epsilon_{\bm{k}-\bm{p}}}+\frac{v_{\bm{p}}^2n_{\bm{p}}}{i\omega_n+E_{\bm p}-\epsilon_{\bm{k}-\bm{p}}}\Bigg).
\end{split}
\end{align}
Note that here we take $\epsilon_B = -1/2 m_r a_{\text{IF}}^2 < 0$.
In order to remove the divergence in both pair propagators, we add and subtract $ \text{Re}[\Pi_v(0,\epsilon_F)] $ in the denominator and with this we can define
\begin{align}
\begin{split}
\frac{1}{g}&= \Pi_v(0,\epsilon_B) - \text{Re}[\Pi_v(0,\epsilon_F)] \simeq -\frac{\pi}{m_r} \frac{1}{\ln \left(k_F a\right)},
\label{gpert}
\end{split}
\end{align}
where we left out $m_r/m$ since $k_Fa_{\text{IF}} \gg m_r/m$.
\section{Perturbative expansion}
\label{appendixB}

Going back to the full  expression for the scattering matrix, we can expand it in the weak coupling regime ($k_Fa_{\text{IF}} \gg 1$) in a perturbative series up to second order in $g$
\begin{align}
\begin{split}
	\mathcal{T}(\bm{k},i\omega_n)	 = \frac{g}{1 - g\Delta\Pi(\bm{k},i\omega_n)} \simeq g + g^2\Delta\Pi(\bm{k},i\omega_n),
\end{split}
\end{align}
where $g$ is expressed by Eq. \eqref{gpert} and
\begin{align}
\begin{split}
	\Delta\Pi(\bm{k},i\omega_n) = \Pi(\bm{k},i\omega_n) -  \text{Re}[\Pi_v(0,\epsilon_F)].
\end{split}
\end{align}
To first order this gives the mean-field contribution to the self-energy of the polaron
\begin{align}
\begin{split}
&\Sigma_1(\bm{q},i\Omega_\nu)=
\vcenter{\hbox{\begin{tikzpicture}
\begin{feynman}
\vertex (a);
\vertex[above=1cm of a] (b);
\diagram* {
(a) --[photon] (b)};
\end{feynman}
\draw[arc arrow={to pos 0.55 with length 2mm}] (b) arc(-90:270:0.5);
\end{tikzpicture}}}
=2g
\sum_{k}
G_{11}(k)
\label{eq:sig0}
\end{split}
\end{align}
with $k = (\bm{k}, i\omega_n)$ and $\sum_{k} = \frac{1}{\beta\mathcal{V}}\sum_{\bm{k}} \sum_{n} $.
After a straightforward calculation we find 
\begin{align}
\begin{split}
	\Sigma_1(0, \epsilon_P) =
	\frac{g}{\pi}\int_{0}^\infty dk\, k
\left(u_{\bm{k}}^2 n_{\bm k}+v_{\bm{k}}^2 (1-n_{\bm k})\right).
\end{split}
\end{align}
The second order contribution is written in Eq. \eqref{Selfenergy} of the main text with
\begin{align}
\chi(k) = \sum_q \Big( G_{11}(q) G_{11}(p-k+q) + G_{12}(q) G_{12}(p-k+q) \Big).
\end{align}
We can write this in a different way as $\Sigma_2=\Sigma_{2a}+\Sigma_{2b}$, where 
\begin{align}
\begin{split}
	&\Sigma_{2a}(0, \epsilon_P) =
\vcenter{\hbox{\begin{tikzpicture}
\begin{feynman}
\vertex (a);
\vertex[right=1cm of a] (a1);
\vertex[right=1.3cm of a1] (a2);
\vertex[above=1.3cm of a1] (c);
\vertex[above=1.18cm of a1] (c1);
\vertex[above=1.3cm of a2] (d);
\vertex[above=1.18cm of a2] (d1);
\vertex[right=1cm of a2] (b2);
\diagram* {
(a1) --[photon] (c1);
(a2) --[photon] (d1);
(a1) --[fermion, edge label'={$k-p,i$}, font=\fontsize{8}{8} ,arrow size=0.14em] (a2);
(c) --[fermion, half left, edge label={$p,\sigma$}, font=\fontsize{8}{8}, arrow size=0.14em] (d),
(d) --[fermion, half left, edge label={$k,\sigma$}, font=\fontsize{8}{8}, arrow size=0.14em] (c)};
\end{feynman}
\end{tikzpicture}}}
\\&=\frac{2g^2}{\beta \mathcal{V}^2}
\sum_{\bm{k},\bm{p}}\sum_n
G_{11}(\bm{k},i\omega_n)
\Delta\Pi(\bm{k},i\omega_n + \epsilon_P).
\raisetag{20pt}
\end{split}
\end{align}
After a straightforward calculation we find
\begin{align}
\begin{split}
	&\Sigma_{2a}(0, \epsilon_P) 
=
\frac{2g^2}{ \mathcal{V}^2}
\sum_{\bm{k},\bm{p}}
\Big(\frac{v_{\bm{p}}^2 u_{\bm{k}}^2 n_{\bm{k}} n_{\bm{p}}}{\epsilon_P + E_{\bm k} +E_{\bm p}-\epsilon_{\bm{k}-\bm{p}}}
\\&+\frac{u_{\bm{p}}^2 v_{\bm{k}}^2 (1-n_{\bm{k}}) (1-n_{\bm{p}})}{\epsilon_P - E_{\bm k}-E_{\bm p}-\epsilon_{\bm{k}-\bm{p}}}
+\frac{v_{\bm{p}}^2 v_{\bm{k}}^2 (1-n_{\bm{k}})  n_{\bm{p}}}{\epsilon_P - E_{\bm k} + E_{\bm p}-\epsilon_{\bm{k}-\bm{p}}}
\\&+\frac{u_{\bm{p}}^2 u_{\bm{k}}^2 n_{\bm{k}} (1-n_{\bm{p}})}{\epsilon_P + E_{\bm k} - E_{\bm p} - \epsilon_{\bm{k}-\bm{p}}}
-\frac{u_{\bm{k}}^2 n_{\bm{k}} + v_{\bm{k}}^2 (1-n_{\bm{k}}) }{\epsilon_F - \epsilon_F(\bm{p}) - \epsilon_I(\bm{p})}\Big).
\raisetag{20pt}
\end{split}
\end{align}
We also get 
\begin{align}
\begin{split}
&\Sigma_{2b}(0,\epsilon_P)=\vcenter{\hbox{\begin{tikzpicture}
\begin{feynman}
\vertex (a);
\vertex[right=1cm of a] (a1);
\vertex[right=1.3cm of a1] (a2);
\vertex[above=1.3cm of a1] (c);
\vertex[above=1.18cm of a1] (c1);
\vertex[above=1.3cm of a2] (d);
\vertex[above=1.18cm of a2] (d1);
\vertex[right=1cm of a2] (b2);
\diagram* {
(a1) --[photon] (c1);
(a2) --[photon] (d1);
(a1) --[fermion, edge label'={$k-p,i$}, font=\fontsize{8}{8} ,arrow size=0.12em] (a2);
(c) --[majorana, half left, edge label={$k,\sigma$}, font=\fontsize{8}{8} ,arrow size=0.12em] (d),
(d) --[anti majorana, half left, edge label={$p,\sigma$}, font=\fontsize{8}{8} ,arrow size=0.12em] (c)};
\end{feynman}
\end{tikzpicture}}}
\\&=\frac{2g^2}{\beta\mathcal{V}}
\sum_{\bm{k}}
\sum_{n} 
G_{12}(\bm{k},i\omega_n)
\Pi_{21}(\bm{k}, i\omega_n + \epsilon_P),
\end{split}
\end{align}
where 
\begin{align}
	\begin{split}
&\Pi_{21}(\bm{k},i\omega_n + \epsilon_P)=\frac{1}{\beta\mathcal{V}}
\sum_{\bm{p}}
\sum_{m} 
G_{12}(\bm{p},i\omega_m)
\\&
G_{0,i}(\bm{k}-\bm{p},i\omega_n-i\omega_m + \epsilon_P).
	\end{split}
\end{align}
After a straightforward calculation we can write
\begin{align}
\begin{split}
	&\Sigma_{2b}(0,\epsilon_P) =\frac{2g^2}{\mathcal{V}^2}\sum_{\bm{k},\bm{p}}u_{\bm{k}}v_{\bm{k}}u_{\bm{p}}v_{\bm{p}}
\Bigg[
\frac{n_{\bm k}n_{\bm p}}{\epsilon_P+E_{\bm k}+E_{\bm{p}}-\epsilon_{\bm{k}-\bm{p}}}
\\&+
\frac{(1-n_{\bm k})(1-n_{\bm p})}{\epsilon_P-E_{\bm{k}}-E_{\bm{p}}-\epsilon_{\bm{k}-\bm{p}}}
-
\frac{(1-n_{\bm k})n_{\bm p}}{\epsilon_P-E_{\bm{k}}+E_{\bm{p}}-\epsilon_{\bm{k}-\bm{p}}}
\\&-
\frac{n_{\bm k}(1-n_{\bm p})}{\epsilon_P+E_{\bm k}-E_{\bm{p}}-\epsilon_{\bm{k}-\bm{p}}}
\Bigg].
\raisetag{17pt}
\end{split}
\end{align}


\bibliography{libraryUsed}
\end{document}